\documentclass{ifacconf}

\usepackage{graphicx}      
\usepackage{natbib}        
\usepackage{float}         
\usepackage{xcolor}        
\usepackage{amsmath}       
\usepackage{amsfonts}      
\usepackage{makecell}      
\definecolor{revgreen}{RGB}{0,140,0}
\definecolor{termblue}{RGB}{0,70,200}
\definecolor{corred}{RGB}{215,0,0}
\renewcommand{\cor}[1]{{\color{corred}#1}}            
\begin{document}
\begin{frontmatter}

\title{Model Augmentation for thermo-mechanical hysteretic dynamics\thanksref{footnoteinfo}} 

\thanks[footnoteinfo]{This project is co-financed by Holland High Tech, top sector High-Tech Systems and Materials, with a PPP innovation subsidy for public-private partnerships for research and development. This work has been submitted to IFAC for possible publication.}

\author[First]{Samarth Toolhally} 
\author[Second]{Ozan Celik} 
\author[Second]{Daniel Deenen}
\author[First]{Maarten Schoukens}
\author[First]{Siep Weiland}

\address[First]{Control Systems group, Eindhoven University of Technology, The
Netherlands (e-mail: s.toolhally@tue.nl, m.schoukens@tue.nl, s.weiland@tue.nl).}
\address[Second]{ASML Netherlands B.V., The
Netherlands (e-mail: ozan.celik@asml.com, daniel.deenen@asml.com)}

\begin{abstract}                
Accurate prediction of thermo-mechanical wafer deformation is essential for sub-nanometer overlay in extreme ultraviolet (EUV) lithography. Localized heating expands the wafer, while friction between the wafer and its supporting elastic pillars causes hysteresis. Modeling this is challenging because of the system’s large scale, the computational cost of friction-induced hysteresis and the sparse deformation measurements. This work augments a linear thermo-mechanical baseline model with a learning-based structure in a linear fractional representation (LFR) framework, identified from sparse measurements. Simulation results show that the augmented model captures the unmodeled dynamics with hysteretic effects and predicts the deformation significantly more accurately than the baseline.
\end{abstract}

\end{frontmatter}

\section{Introduction}
Photolithography demands extremely high precision: during wafer exposure, the exposed region must be accurately aligned with the projected image to ensure the functionality and yield of the fabricated chips \citep{DavidvandenHurk}, and this requirement tightens as feature sizes continue to shrink following Moore's law.

\begin{figure}[htbp]
    \centering
    \includegraphics[scale=0.5]{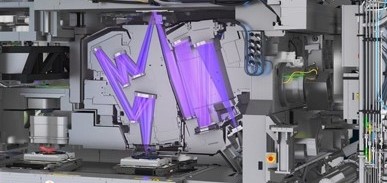}
    \caption{Wafer stage of a lithography machine.}
    \label{asml machine}
\end{figure}
In extreme ultraviolet (EUV) wafer scanners, the light is not projected onto the entire field at once, but focused on a smaller region, the slit, which moves at constant velocity along the field. The high EUV source power induces significant thermal loads on the wafer. Since semiconductor devices consist of multiple stacked layers that must be precisely aligned, the resulting deformation causes overlay errors: heat-induced wafer deformation alone can result in overlay errors of up to 4 nm \citep{54e23470bab2456ebc98e4c620a3314b2}, while the overall overlay error target is much lower.

\begin{figure}[htbp]
    \centering
    \includegraphics[scale=0.4]{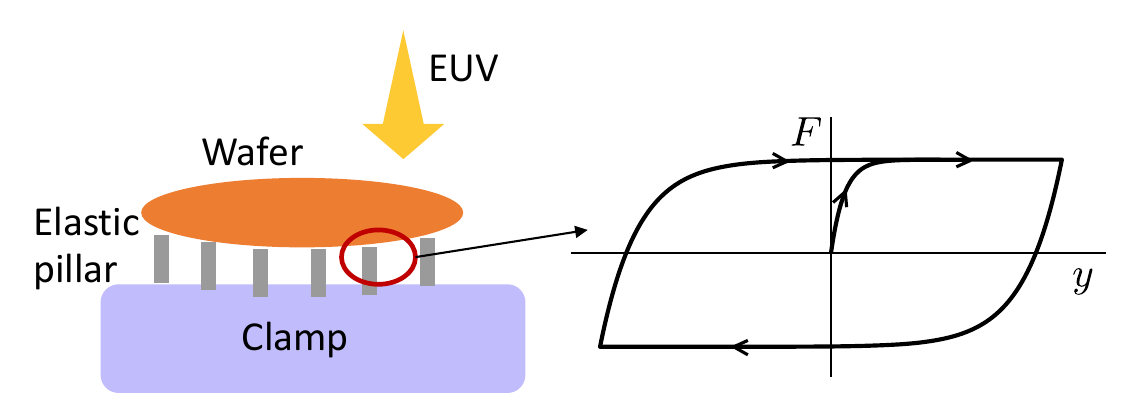}
    \caption{Wafer clamp unit, consisting of wafer and elastic pillars.}
    \label{wafer}
\end{figure}

The wafer is secured by a clamp containing several thousand elastic pillars (Fig.~\ref{wafer}), which minimize the contact area and thereby reduce adhesion and ease wafer removal. During exposure, the wafer stage follows a meandering trajectory, and the heat-induced expansion of the wafer interacts with the pillars through frictional contact, which makes the deformation hysteretic.

\subsection{Modeling Challenges and Approaches}
Three properties make this deformation difficult to model. First, during exposure, no sensors can be used to measure the deformation, so real-time overlay compensation requires a predictive model that is both accurate and computationally efficient. Second, deformation measurements for building such a model are sparse, i.e., markers on the wafer only provide deformation information within the exposed slit region, so the identification method must learn from partial observations of the deformation field. Third, the deformation is hysteretic, i.e., the output depends not only on the current input but also on past inputs and internal states, so the same input may produce different outputs depending on the system's history \citep{1064347}, which complicates model development and parameter estimation.

First-principles (FP) models \citep{LjungGlad1994} are derived from physical laws and are therefore interpretable. Several methods capture static and dynamic hysteresis \citep{Schuderer20252,KRIKELIS2021702}. Here, however, the hysteresis of thousands of pillars must be represented, which makes such models computationally expensive. On the other hand, black-box models such as artificial neural networks (ANNs) are universal function approximators \citep{10.5555/70405.70408, SJOBERG19951691} that require little modeling effort and can be highly accurate, but they lack physical interpretability and may generalize poorly outside the training conditions. Hybrid approaches combine both. Physics-informed learning embeds the governing equations as a soft constraint in the training loss \citep{Karniadakis2021}, whereas model augmentation retains a baseline model in the model structure and learns only the unmodeled dynamics \citep{Bolderman20242,HOEKSTRA2025101304}. If the baseline already captures the dominant behavior, model augmentation can improve accuracy while keeping the computational efficiency required for real-time use.

\subsection{Contribution}
This work investigates model augmentation for predicting friction-induced hysteretic wafer deformation during EUV exposure. The contributions are:
\begin{itemize}
  \item An LFR-based augmentation of a linear thermo-mechanical baseline
        model, in which additional latent states carry the slip memory
        that the baseline lacks, and neural networks correct the state
        update and the output;
  \item Identification from sparse measurements using SUBNET: an encoder
        estimates the initial state from a short window of past heat
        inputs, heating locations and measured deformations, so full
        state measurements are not required, and a masked multi-step
        loss ensures that only the measured (heated) pillars contribute
        to the training objective;
  \item A simulation study on a 576-pillar wafer-clamp model, showing that
        the augmented model generalizes to unseen heat loads with a
        $17$-fold lower free-run simulation error than the baseline, while
        simulating about $546$ times faster than the full model.
\end{itemize}
Section~\ref{sec:system} describes the thermo-mechanical system, Section~\ref{sec:aug} the augmentation structure and identification method, Section~\ref{sec:results} the simulation results, and Section~\ref{sec:concl} concludes.

\section{Thermo-Mechanical System}\label{sec:system}

Exposure to the EUV source induces thermal expansion of the wafer. At the same time, the wafer interacts mechanically with the elastic pillars through frictional contact. Initially, thermal expansion causes the elastic pillars to deform elastically. Once the thermally induced force exceeds the maximum static friction force\cor{,} relative motion occurs between the wafer and the elastic pillars, and a nearly constant friction force acts at the contact interface. The interaction between thermal, frictional and mechanical forces gives rise to friction-induced hysteresis.

The dynamics of the true system are of the form
\begin{equation}
\begin{split}
    x_{k+1} &= f(x_k,u_k),\\
    y_k &= h(x_k,u_k).
\end{split}\label{data sys}
\end{equation}
Here $x_k \in \mathbb{R}^{n_x}$ consists of the temperatures and slip states of the elastic pillars, defined on a finite set of grid points covering the wafer. The input $u_k \in \mathbb{R}^{n_u}$ denotes the heat load generated by the EUV source, which follows a meandering scanning pattern. The output $y_k \in \mathbb{R}^{n_y}$ stacks the wafer deformations at all pillars in both in-plane directions, measured relative to their undeformed positions.

\begin{figure}[htbp]
    \centering
    \includegraphics[scale=0.21]{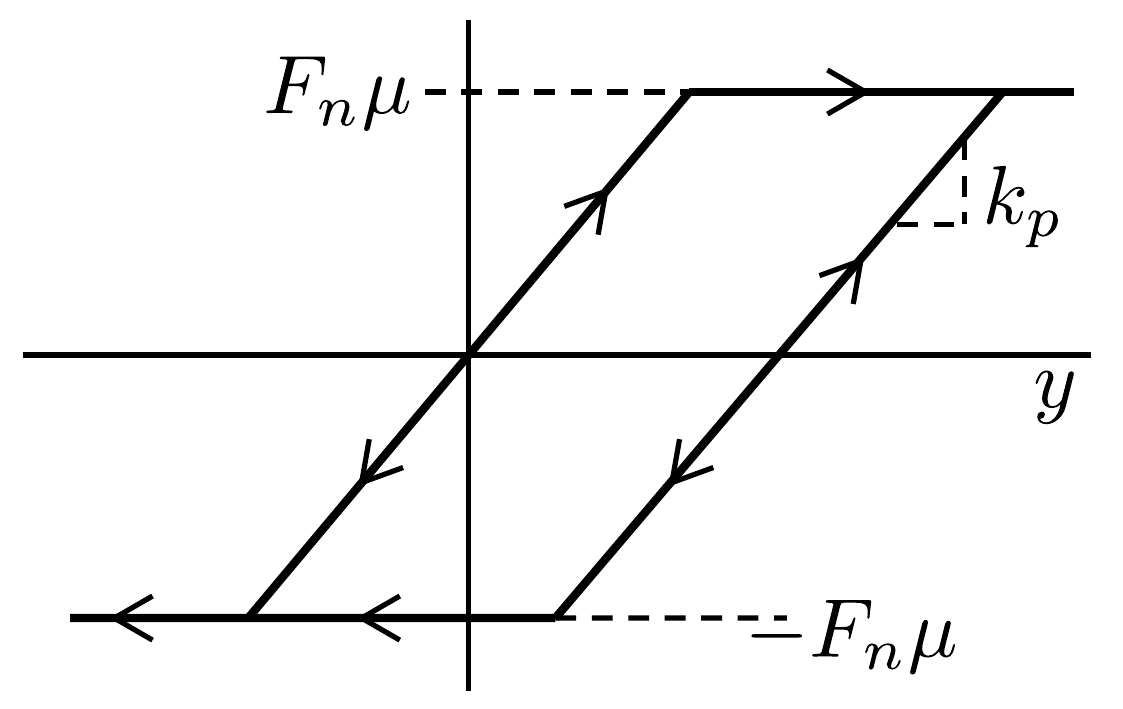}
    \caption{Force-deformation hysteresis loop of a 1-D Jenkins element. Linear elastic behavior is observed during sticking, while force saturation at the friction threshold produces hysteretic stick-slip dynamics.}
    \label{jenkis}
\end{figure}

The function $f:\mathbb{R}^{n_x}\times\mathbb{R}^{n_u}\rightarrow\mathbb{R}^{n_x}$ describes the state transition dynamics, whereas $h:\mathbb{R}^{n_x}\times\mathbb{R}^{n_u}\rightarrow\mathbb{R}^{n_y}$ maps the states and inputs to the wafer deformation. The dominant nonlinearity in the system originates from the stick-slip interaction between the wafer and the elastic pillars. An elastic pillar either sticks to the wafer surface or slips relative to it, depending on the local wafer deformation induced by thermal loading. To capture this behavior, the Maxwell-slip (MS) model is employed, which is widely used to describe frictional hysteresis and slip dynamics \citep{Ruderman2011,Qin2024}. The MS model uses Jenkins elements to describe slip and friction dynamics. A Jenkins element consists of a linear spring with Coulomb friction. A Jenkins element is defined for each elastic pillar, and the hysteresis loop for a 1-D Jenkins element is shown in Fig.~\ref{jenkis}. This stick-slip condition is given by

\begin{equation}\label{eq:jenkins}
  F = \begin{cases}
        k_p\,\bar{y} & \text{if stick,}\\
        F_n \mu, & \text{otherwise},
      \end{cases}
\end{equation}
where $k_p$ is the spring constant, $\mu$ is the coefficient of friction, $F_n$ is the normal force acting on the wafer due to its weight, and $\bar{y}$ is the elongation of the spring relative to its friction contact.

The transition between the stick and slip regimes depends on the deformation state of the wafer, introducing a nonlinear hysteretic behavior in the input-output relationship. As a consequence, the output mapping becomes implicit, since the current deformation and friction force must satisfy the stick-slip conditions simultaneously. Determining the output thus requires solving the equations numerically, which significantly increases the computational cost for predicting the deformation.

\begin{figure}[htbp] \centering \includegraphics[scale=0.25]{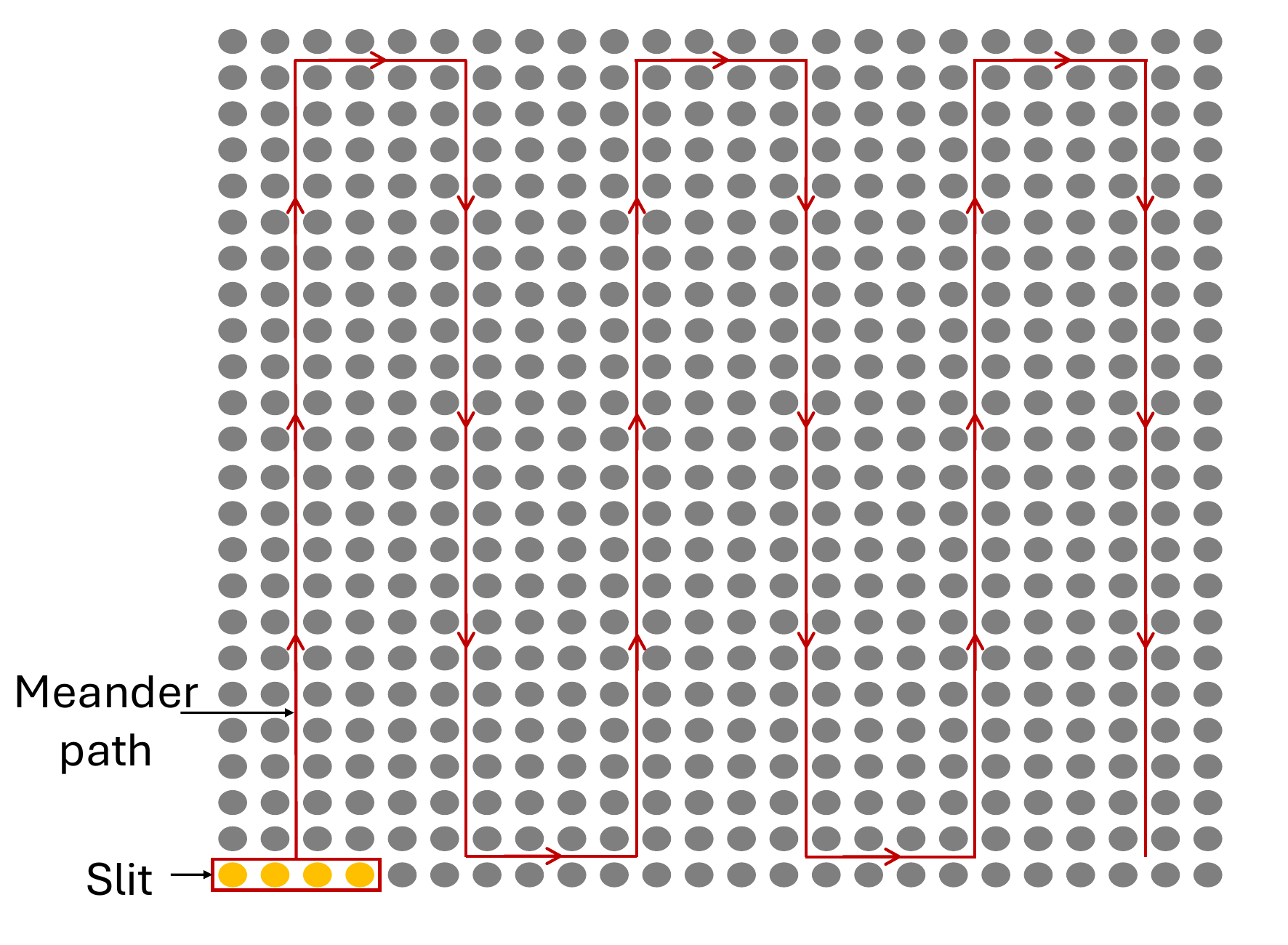} \caption{Elastic pillar arrangement showing the meandering path for EUV exposure.} \label{elastic pillarexposer} \end{figure}

In the setup, the wafer is discretized into a $24 \times 24$ grid of elastic pillars, each modeled as a Jenkins element and coupled to a thermal model of the wafer. The EUV slit covers $4$ adjacent elastic pillars and advances one row per time step along a meander of six lanes of 4 pillars width. One full exposure therefore takes $144$ steps, and each elastic pillar is heated exactly once by the EUV. 
The input $u_k \in \mathbb{R}^{576}$ is the vector of heat loads applied to all pillars at time $k$, and the output $y_k \in \mathbb{R}^{1152}$ is the vector of in-plane deformations of all pillars at time $k$.
The heat source follows a meandering exposure path from one edge of the wafer to the other, as illustrated in Fig.~\ref{elastic pillarexposer}. This pattern qualitatively captures the key characteristics of a real scanning exposure. At each time instant, deformation measurements are recorded only for the four actively heated elastic pillars, while no measurements are available for the remaining regions. Consequently, the output measurements are highly sparse. This is described by the binary mask $H_k \in \{0,1\}^{n_y}$, whose entries are $1$ for the in-plane outputs of the heated pillars at time $k$ and $0$ otherwise, and by the measurement model,
\begin{equation}\label{eq:meas}
  y'_k = S_k\, y_k + e_k,
\end{equation}
where $S_k \in \{0,1\}^{n_{y'} \times n_y}$ selects the entries of $y_k$ for which $H_k = 1$,
$y'_k \in \mathbb{R}^{8}$ is the in-plane deformation of the 4 heated
pillars, and $e_k$ is white Gaussian measurement noise, specified in Section~\ref{sec:results}. The other entries of $y_k$ are not measured.

\section{Model Augmentation}\label{sec:aug}
The goal is to obtain a faster surrogate of model \eqref{data sys}. To achieve this, the linear-fractional representation (LFR)-based
model augmentation approach, as developed in \citep{HOEKSTRA2025101304} is used. This method identifies missing system dynamics by leveraging an artificial neural network (ANN) parameterization for the learning component and prior information in the baseline model, and offers flexibility in accommodating a wide range of model augmentation structures.

For our study, we use \eqref{data sys} as the data generating system, and use a linear state-space model as the baseline.\\
To obtain the linear baseline model, we ignore the slip part of the hysteretic dynamics, i.e., we consider only the stick regime, hence it behaves like a linear spring model. The resulting state-space equation is given by
\begin{subequations}\label{eq:baseline}
\begin{align}
  \tilde{x}_{k+1} &= A \tilde{x}_k + B u_k, \\
  \tilde{y}_k &= C \tilde{x}_k+Du_k.
\end{align}
\end{subequations}
Here, the first equation represents the thermal state evolution and the second gives the elastic pillar deformation for the linear model. The thermal part of the baseline is exact, so its prediction error
originates only from neglecting slip. Model augmentation is used to
capture this residual between \eqref{data sys} and \eqref{eq:baseline}.
\subsection{Model augmentation structure}
\begin{figure}[htbp]
    \centering
    \includegraphics[scale=0.4]{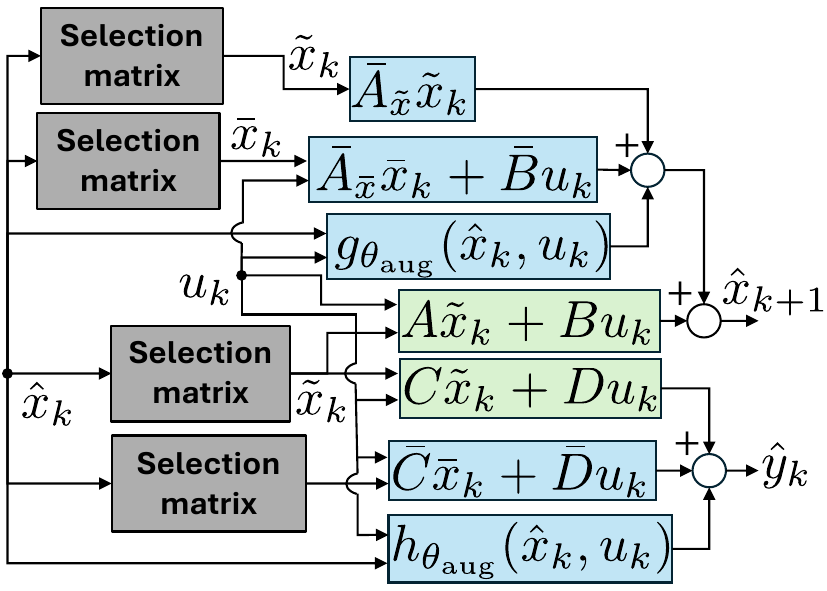}
    \caption{LFR-based model augmentation structure. A linear baseline model is augmented with additional latent states and neural network based correction terms to capture unmodeled hysteretic dynamics.}
    \label{aug model}
\end{figure}


The baseline model \eqref{eq:baseline} describes the thermal dynamics
exactly and the deformation correctly as long as all pillars stick;
what it misses is the slip states\cor{,} which store the loading history. The augmentation therefore adds two components to the
baseline while keeping it intact: (i) latent states $\bar{x}_k$ that
carry the slip information, and (ii) neural networks that correct the state
update and the output where slip occurs. The resulting model state is
$\hat{x}_k = [\,\tilde{x}_k;\; \bar{x}_k\,] \in \mathbb{R}^{n_{\hat{x}}}$,
with $\tilde{x}_k$ the baseline states, and $\hat{y}_k \in
\mathbb{R}^{n_y}$ is the model output. The structure is shown in Fig.~\ref{aug model}.

Learning the full dynamics with a black-box network would require relearning the known linear part as well. Instead, the networks only model the residual, which has been reported to improve data efficiency, accelerate training, and lead to better generalization performance\cor{~}\citep{9368834,9655207}. The resulting parameterization of the augmented state-transition and output equations is given by

\begin{equation}\label{aug equation}
\begin{split}
  \begin{bmatrix}
            \tilde{x}_{k+1} \\
        \bar{x}_{k+1}
\end{bmatrix}  &=   \begin{bmatrix}
            A &0 \\
        \bar{A}_{\tilde{x}} & \bar{A}_{\bar{x}}
\end{bmatrix}   \begin{bmatrix}
            \tilde{x}_{k} \\
        \bar{x}_{k}
\end{bmatrix} +   \begin{bmatrix}
            B \\
        \bar{B}
\end{bmatrix} u_k +   \begin{bmatrix}
            0 \\
        g_{\theta_{\text{aug}}}(\hat{x}_k,u_k)
\end{bmatrix},\\
\hat{y}_k &=
    \begin{bmatrix}
        C &\bar{C}
    \end{bmatrix}
    \begin{bmatrix}
        \tilde{x}_k\\
        \bar{x}_k
    \end{bmatrix}+
h_{\theta_{\text{aug}}}(\hat{x}_k,u_k) + 
        Du_k+ \bar{D}
u_k.
\end{split}
\end{equation}
Here, the $\bar{A}_{\tilde{x}}$, $\bar{A}_{\bar{x}}$, $\bar{B}$, $\bar{C}$ and $\bar{D}$ are parameterized matrices of appropriate dimensions while $g_{\theta_{\text{aug}}}(\hat{x}_k,u_k)$ and $h_{\theta_{\text{aug}}}(\hat{x}_k,u_k)$ are feedforward neural networks. All the parameters of the linear augmentation terms and the neural networks are stacked in  $\theta_{\text{aug}}\in \mathbb{R}^{n_{\theta_{\text{aug}}}}$. The linear augmentation terms provide additional flexibility to capture structured dynamics that are not represented by the baseline model, whereas the neural networks account for the remaining nonlinear effects not captured by the baseline model. The baseline matrices $A$, $B$, $C$ and $D$ remain fixed during identification.
\subsection{Identification method}
\begin{figure}[htbp]
    \centering
    \includegraphics[scale=0.4]{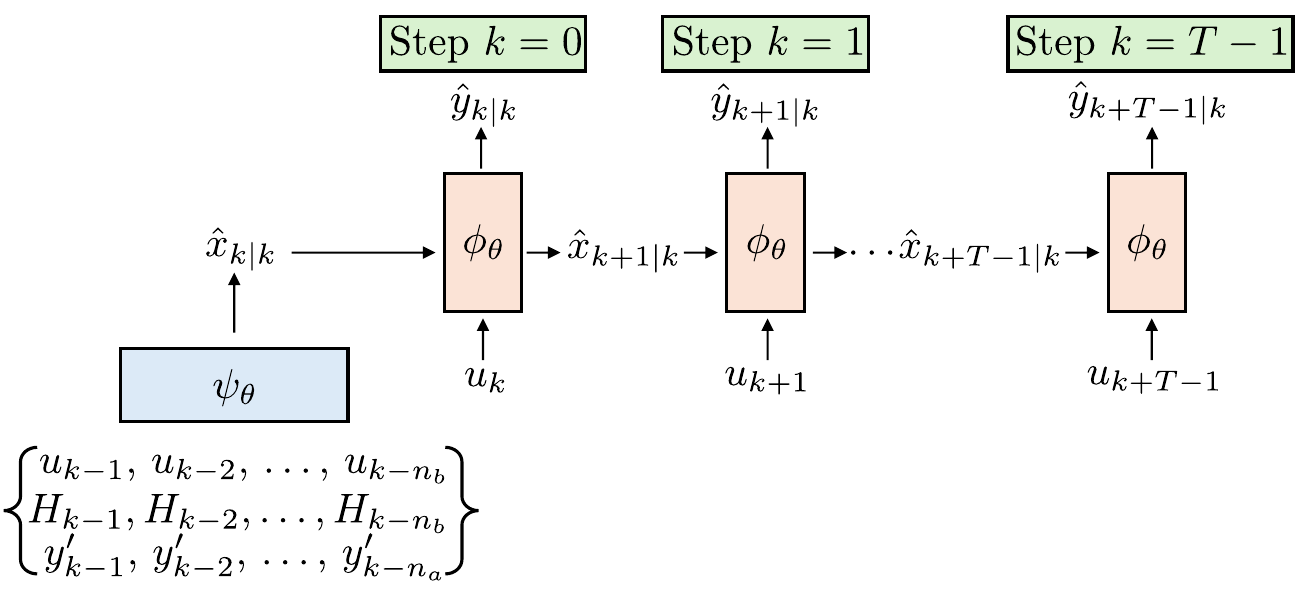}
    \caption{The encoder \(\psi_{\theta}\) estimates the initial state at time \(k\) using past deformation measurements, heat inputs, and heating locations, the estimated state is then propagated through \(\phi_{\theta}\) over a prediction horizon of length \(T\).}
    \label{encoder}
\end{figure}

For identification of the augmented system \eqref{aug equation}, the \textit{subspace encoder network} (SUBNET) method is used \citep{Beintema2023,pmlr-v144-beintema21a}, which provides statistically consistent estimates of nonlinear state-space (NL-SS) models and has demonstrated state-of-the-art performance in several benchmark datasets. 
To enable computational efficiency, it employs a truncated objective function that facilitates the use of batch optimization methods commonly adopted in machine learning. This objective is constructed by dividing the available data into multiple segments of length $T$ from the available data $D_N = \{(y_k,u_k)\}^N_{k=1}$ obtained from the data-generating system \eqref{data sys} as shown in Fig. \ref {encoder}.
The loss function is defined as 
\begin{subequations}
    \begin{align}
        V(\theta) = \frac{1}{2N(T+1)} \sum_{i=1}^{N} \sum_{\ell=0}^{T} \|e_{i,\ell}\|_2^2,\\
        \begin{bmatrix}
        \hat{x}_{k_i|k_i+\ell+1} \\
        \hat{y}_{k_i|k_i+\ell}
    \end{bmatrix}
    := \phi_\theta \left( \hat{x}_{k_i|k_i+\ell}, u_{k_i+\ell} \right), \\
    \hat{x}_{k|k} := \psi_\theta \left( y'_{k-n_a:k-1}, u_{k-n_b:k-1}, H_{k-n_b:k-1} \right).
    \end{align}
\end{subequations}
Here
\begin{equation} e_{i,\ell} = H_{k_i+\ell} \odot \left( \hat{y}_{k_i|k_i+\ell} -y_{k_i+\ell} \right), \end{equation}
and $H_k \in \{0,1\}^{n_y}$ is the indicator of the heated elastic pillars at time $k$ defined in Section~\ref{sec:system}, $\odot$ denotes element-wise multiplication. Here $\theta = [\theta_{\text{aug}}\; \theta_{\text{encoder}}]$ are the augmented-model and encoder parameters and $\phi_{\theta}$ is the augmentation structure \eqref{aug equation}. $k|k+\ell$ indicates the state $\hat{x}_{k}$ or output $\hat{y}_{k}$ at time $k+\ell$ simulated from the initial state $\hat{x}_{k|k}$ at time $k$. The subsections start from a randomly selected time $k \in \{n+1, \dots, N - T\}$, $n=\text{max}(n_a,n_b)$. 

The model is trained using a masked multi-step prediction error criterion. For each training window, the encoder first estimates the latent state.  To estimate this state, the encoder uses past input-output and heated index information. To reduce the complexity of the encoder, only the deformation information of heated elastic pillars is used. Hence $y_k \in \mathbb{R}^{n_y}$ is compressed to $y_k' \in \mathbb{R}^{n_{y'}}$, where $y_k'$ consists of the deformation of only the heated elastic pillars at time instance $k$. Because $y'_k$ always contains the four currently heated pillars, its spatial meaning changes over time; the heating locations $H_{k-n_b:k-1}$ are therefore also passed to the encoder, so that it can associate each measurement with its position on the wafer. This encoder is co-estimated with $\phi_\theta$.

The identified state-space model is then simulated forward over a prediction horizon $T$, and the predicted deformation field is compared with the measured deformation field to compute the loss. Only prediction errors at heated and measured locations contribute to the optimization objective, while unobserved grid cells are ignored, using masking. This masking operation directly affects the gradient, i.e., measured cells produce non-zero gradients that update the encoder and model parameters, whereas unmeasured cells yield zero gradients and therefore have no influence on the parameter updates. As a result, the encoder, state-transition model, and output model are trained exclusively using information from the available deformation measurements, ensuring consistency with the physical sensing configuration where deformation is only observed at heated locations.
\subsection{Initialization}
A good initialization of the model parameters is important so that the nonlinear optimization does not converge to a poor local minimum. With random initialization, the initial predictions can be highly inaccurate, causing the nonlinear optimization algorithm to search in an unfavorable region of the parameter space. Because the training objective is typically non-convex, the optimizer may converge to a poor local minimum, leading to slow convergence, unstable training, or reduced model accuracy.  To mitigate these issues, an improved initialization strategy based on the baseline linear model is adopted \citep{9655207,SCHOUKENS2020310}. The key idea is to incorporate prior knowledge of the system dynamics into the neural state-space model, thereby providing the optimizer with a physically meaningful starting point.
Since $g_{\theta_{\text{aug}}}(\cdot)$ and $h_{\theta_{\text{aug}}}(\cdot)$ are feedforward neural networks, we can write them as 
\begin{equation}
\begin{split}
     g_{\theta_{\text{aug}}}(\hat{x}_k,u_k) = W^{g}_{\text{last}}\Omega^{g}(\hat{x}_k,u_k)+b^{g}_{\text{last}},\\
     h_{\theta_{\text{aug}}}(\hat{x}_k,u_k) = W^{h}_{\text{last}}\Omega^{h}(\hat{x}_k,u_k)+b^{h}_{\text{last}}.
\end{split}
\end{equation}
Here, $\Omega$ indicates the output of the last hidden layer after activation. Now, initialising $\bar{A}_{\tilde{x}}$, $\bar{A}_{\bar{x}}$, $\bar{B}$, $\bar{C}$, $\bar{D}$, $W_{\text{last}}$ and $b_{\text{last}}$ to zero, the initial model behaves like the baseline model. All other ANN layer parameters are initialized randomly using uniform distribution between $-1$ and $1$, $\mathcal{U}(-1,1)$. The initialization assumes that the input-output data are normalized to zero mean and unit standard deviation; such normalization improves the conditioning of the optimization problem and speeds up training \citep{Bishop1995NeuralNF}.

\section{Simulation results}\label{sec:results}
\subsection{Setup}
 To obtain datasets that differ in more than the noise realization, the scan starts from the four wafer corners, runs along either of the two in-plane directions, and uses one of five heat-load levels, $0.6, 0.8, \dots, 1.4$ times the nominal power. On top of this, every exposure step is perturbed by white Gaussian noise of variance $0.01$,i.e., a standard deviation of $10\%$ of the nominal power. Each dataset contains $144$ time samples. The models are trained on $1200$ datasets and validated on $400$ datasets with new noise realizations. The models are tested on $32$ datasets with heat-load levels $0.7, 0.9, 1.1$ and $1.3$ times the nominal power, which lie between the training levels and were never used for training, and with an input noise variance of $0.03$, three times that of the training data (standard deviation about $17\%$ of the nominal power). The simulated deformations are noise-free; to emulate sensor noise, white Gaussian noise with a standard deviation of $2\%$ of the RMS deformation at nominal power is added to all measured outputs $y'_k$ used for training and validation, while the test error is computed with respect to the noise-free deformation.

The predictive performance of the identified models is evaluated using the normalized root mean square error (NRMSE) between the true and predicted deformations. Since the objective of the augmented model is to estimate the deformation only at the actively heated elastic pillars, the performance metric is computed only using these measurements. The RMSE and NRMSE are defined as 
\begin{equation} \begin{split} \mathrm{RMSE} &= \sqrt{\frac{1}{n_{y'}N}\sum_{k=1}^{N}\left\|y'_k-\hat{y}'_k\right\|_2^2},\\ \mathrm{NRMSE} &= \frac{\mathrm{RMSE}}{\sigma(y')}, \end{split} \end{equation} 
where \(y'_k\) and \(\hat{y}'_k\) denote the true and predicted deformation of the heated elastic pillars at time instant \(k\), respectively, \(N\) is the total number of samples, and \(\sigma(y')\) denotes the standard deviation of the measured deformation data.\\
Table~\ref{hyperparamtes} summarizes the hyperparameters used for identifying the augmented model. The hyperparameters were chosen based on the structure of the system and the available data. The encoder lag $n_a = n_b = 10$ was selected because the deformation at the heated elastic pillars is mainly determined by the thermal and slip states of recently heated neighboring elastic pillars. Since the slit advances one row per time step, a window of 10 samples covers the deformation and heat inputs of 40 recently exposed elastic pillars, which provides sufficient information to estimate the initial state. Larger lags were avoided because each additional lag increases the encoder input dimension and a larger lag reduces the number of admissible start times $k \in \{n+1,\dots,N-T\}$ for the training subsections, given the short exposure length $N = 144$. For $g_{\theta_{\text{aug}}}$ and $h_{\theta_{\text{aug}}}$, two hidden layers with 32 neurons each are used. A small network was chosen because the linear baseline already captures the dominant thermo-mechanical dynamics, and the memory of the hysteresis is carried by the latent states $\bar{x}_k$.
\begin{table}[htbp] \centering \caption{Hyperparameters used in identification of the augmented model.} \begin{tabular}{c c c c c c} \makecell{Hidden \\ layers} & Neurons & $n_a$, $n_b$ & $T$ & Epochs & \makecell{Batch \\ size} \\ [0.5ex] \hline $2$ & $32$ & $10,10$ & $100$--$130$ & $240$ & $256$ \\ [1ex] \hline \end{tabular} \label{hyperparamtes} \end{table}
\subsection{Model selection}
At each time $k$, the true system contains $1152$ slip states, one for each pillar and in-plane direction, and evaluating them requires solving the stick-slip conditions numerically at every time step. Instead of reproducing these states one by one,
a small number $n_{\bar{x}}$ of latent states is used, under the
assumption that the slip pattern evolves in a structured, spatially
correlated way along the scan and can therefore be summarized by far
fewer degrees of freedom.
 The mappings $\bar{B}$ and $\bar{C}$ are learned from data and determine how the latent states relate to the individual elastic pillars. Increasing $n_{\bar{x}}$ further increases the number of parameters in $\bar{A}$, $\bar{B}$, $\bar{C}$ and the network layers. This raises the risk of overfitting and reduces the computational advantage of the surrogate. To quantify this trade-off, augmented models with $n_{\bar{x}} \in \{25, 50, 100, 200\}$ latent states were identified with otherwise identical settings; their test errors and simulation times are listed in Table~\ref{tab:model_comparison}. Increasing $n_{\bar{x}}$ from $25$ to $100$ reduces the test NRMSE from $0.0386$ to $0.0260$, whereas $200$ latent states give no further improvement ($0.0272$) and only increase the simulation time, from $0.705$\,s to $0.814$\,s. The improvement thus saturates around $100$ latent states, which is far fewer than the $1152$ slip states of the true system.

\begin{figure}[htbp] \centering \includegraphics[scale=0.9]{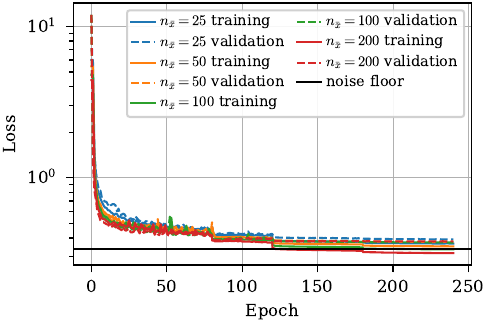} \caption{Training (solid) and validation (dashed) loss of the augmented models with $n_{\bar{x}} \in \{25, 50, 100, 200\}$ latent states. The black line is the noise floor set by the $2\%$ measurement noise.}\label{loss} \end{figure}

Fig.~\ref{loss} shows the training and validation losses of the four augmented models. After epoch $180$ the losses no longer change, indicating convergence. The validation losses settle just above the noise floor, i.e.\ the loss level caused by the added $2\%$ measurement noise, so the models explain nearly all of the deformation that can be explained from the noisy data; for $n_{\bar{x}} \geq 50$ the validation losses are almost equal. The validation loss does not increase towards the end of training for any model size, so no overfitting is visible in the validation loss. However, the training loss of the model with $n_{\bar{x}} = 200$ falls slightly below the noise floor, which means that this model starts to fit the measurement noise; this is consistent with its slightly higher test error compared with $n_{\bar{x}} = 100$ in Table~\ref{tab:model_comparison}.
Also, to check the sensitivity to the network width, the model with $n_{\bar{x}}=100$ was trained with 16 and 64 neurons per hidden layer in $g_{\theta_{\text{aug}}}$ and $h_{\theta_{\text{aug}}}$, with otherwise identical settings. The results are compared in Table~\ref{tab:model_comparison}, which shows similar accuracy and simulation time.
\subsection{Prediction accuracy and computation}
Table~\ref{tab:model_comparison} presents the normalized root mean square error (NRMSE) of the baseline and augmented models on the test datasets. With $n_{\bar{x}} = 100$, the augmented model achieves an NRMSE of $0.0260$, against $0.4351$ for the baseline model, i.e.\ the prediction error is reduced by a factor of approximately $17$. Since the test datasets use heat-load levels and an input noise level that were not seen during training, this shows that the augmented model generalizes within the range of operating conditions covered by the training data.

\begin{table}[htbp]
    \centering
    \caption{Test NRMSE and average simulation time per test dataset for the full model, the baseline, and augmented models with $n_{\bar{x}}$ latent states and hidden layers of the given width.}
    \label{tab:model_comparison}
    \begin{tabular}{lccc}
        Model & Neurons & Test NRMSE & Simulation time (s) \\
        \hline
        Full model           & --  & --     & 385.19 \\
        Baseline             & --  & 0.4351 & --     \\
        $n_{\bar{x}} = 25$   & 32  & 0.0386 & 0.651  \\
        $n_{\bar{x}} = 50$   & 32  & 0.0303 & 0.693  \\
        $n_{\bar{x}} = 100$  & 16  & 0.0279    & 0.701    \\
        $n_{\bar{x}} = 100$  & 32  & \textbf{0.0260} & 0.705 \\
        $n_{\bar{x}} = 100$  & 64  & 0.0274    & 0.713    \\
        $n_{\bar{x}} = 200$  & 32  & 0.0272 & 0.814  \\
        \hline
    \end{tabular}
\end{table}

\begin{figure}[htbp] \centering \includegraphics[scale=0.372]{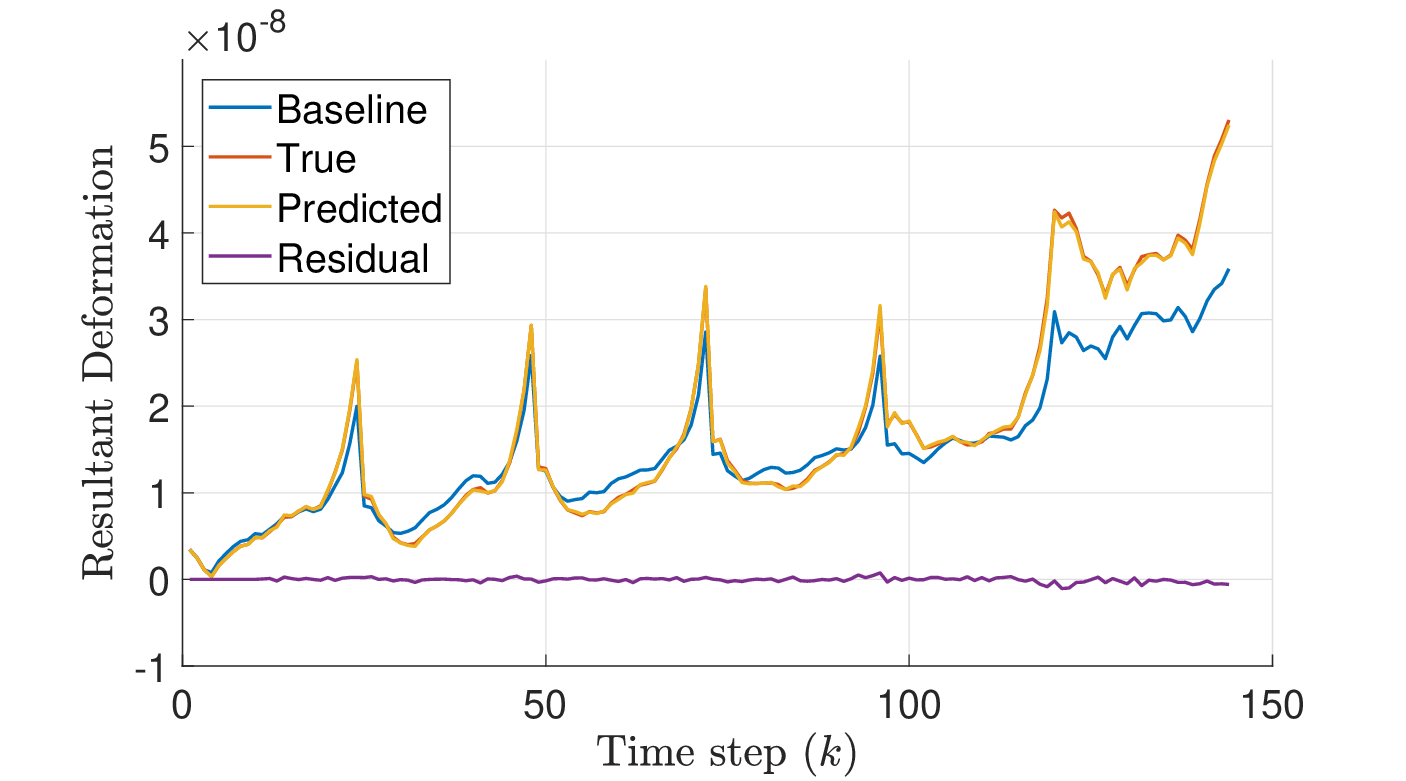} \caption{Resultant deformation of the first pillar in the slit over one exposure of a test dataset: baseline model, true system,  augmented model ($n_{\bar{x}} = 100$), and the residual of the augmented model.} \label{deformationplot} \end{figure}

Fig.~\ref{deformationplot} shows the resultant deformation of the first pillar
in the slit over one exposure of a test dataset, together with the predictions of the baseline and augmented models. The peaks occur at the lane ends, $k = 24, 48, \dots, 120$, where the slit sits in the region of largest deformation. At
the next step, the output jumps to pillars in the adjacent lane, where the deformation is
much smaller, hence we see a dip.
During the first time steps the baseline follows the true deformation, because the thermal load is still small and all pillars stick. Once pillars start to slip, the baseline underestimates the peaks, by up to about $20\%$, and its error grows over the exposure, up to about $30\%$ in the last lane ($k \approx 120$--$144$). This agrees with the physics as the baseline assumes that every pillar sticks and therefore holds the wafer too stiffly, whereas slipping pillars transmit at most the friction force  $F_n\mu$ and keep their shifted position, allowing the wafer to expand further. The residual stays below about $2\cdot 10^{-9}$, i.e.\ a few percent of the peak deformation, and is slightly larger around the turn of the meander path near $k \approx 96$ and in the last lane, where the accumulated slip is largest. At the turns, the spatial distribution of thermal loading changes more rapidly than during straight-line scanning, resulting in localized transient deformation dynamics. Since friction-induced hysteresis depends on the evolution of the underlying slip states, such transients are inherently more difficult to approximate than the smoother dynamics observed during straight-line scanning. Despite these localized prediction errors, the residual does not grow systematically over time, indicating that the model maintains its predictive capability over the entire exposure sequence. Overall, the results demonstrate that the augmented model captures the effect of the friction-induced hysteresis that the baseline model omits, including the deformation accumulated through slip over the exposure, which substantially improves the prediction accuracy under sparse measurement conditions.

To assess the computational efficiency of the proposed augmented model, simulation times were compared with those of the true system. All simulations were performed on a computer equipped with an Intel\textsuperscript{\textregistered} Core\texttrademark{} i7-13700H processor (13th generation) and 16 GB of RAM. The computational time was evaluated over $32$ independent simulations, and the average simulation time was reported. For the true model, implemented in MATLAB R2022b, the average simulation time was approximately $385.19$\,s per run. In contrast, the augmented model with $n_{\bar{x}} = 100$, implemented using deepSI in Python 3.9.13, required only $0.705$\,s per simulation on average. This corresponds to a speed-up factor of approximately $546$. The results indicate that the proposed augmented model not only provides accurate deformation predictions but also offers a significant improvement in computational efficiency.

\section{Conclusion}\label{sec:concl}
This work presented a model augmentation approach for friction-induced hysteretic wafer deformation, identified from sparse, location-varying measurements. A linear baseline is kept intact, latent states carry the slip memory, and an encoder with a masked multi-step loss enables identification from the heated pillars alone, starting from a baseline-based parameter initialization. In a simulation study on a 576-pillar model, the augmented model reached a test NRMSE of $0.0260$ against $0.4351$ for the baseline, a 17-fold reduction, while simulating about 546 times faster than the full model. These results are limited to simulation data with an exact thermal baseline, to heat loads between the training levels, and to evaluation at the heated pillars. Future work includes a formal analysis of data informativity, i.e.\ which slip states can be identified from sparse, location-varying measurements and how the heat load and scan path should be designed for this, as well as validation on measurement data.
\section*{DECLARATION OF GENERATIVE AI AND AI-ASSISTED TECHNOLOGIES IN THE WRITING PROCESS}
During the preparation of this work, the author(s) used Microsoft 365 Copilot in order to assist with textual and grammatical editing. 
After using this tool/service, the author(s) reviewed and edited the content as needed and take(s) full responsibility for the content of the publication.

\bibliography{references}             

@article{Beintema2023,
   author = {Gerben I. Beintema and Maarten Schoukens and Roland Tóth},
   issn = {00051098},
   journal = {Automatica},
   month = {10},
   publisher = {Elsevier Ltd},
   title = {Deep subspace encoders for nonlinear system identification},
   volume = {156},
   year = {2023}
}

@InProceedings{pmlr-v144-beintema21a,
  title = 	 {Nonlinear state-space identification using deep encoder networks},
  author =       {Beintema, Gerben and Toth, Roland and Schoukens, Maarten},
  booktitle = 	 {Proceedings of the 3rd Conference on Learning for Dynamics and Control},
  pages = 	 {241--250},
  year = 	 {2021},
  volume = 	 {144},
  series = 	 {Proceedings of Machine Learning Research},
  month = 	 {07 -- 08 June}
}

@INPROCEEDINGS{9655207,
  author={Schoukens, Maarten},
  booktitle={2021 European Control Conference (ECC)}, 
  title={Improved Initialization of State-Space Artificial Neural Networks}, 
  year={2021},
  volume={},
  number={},
  pages={1913-1918}}

@INPROCEEDINGS{9368834,
  author={Forgione, Marco and Piga, Dario},
  booktitle={2020 IEEE 14th International Conference on Application of Information and Communication Technologies (AICT)}, 
  title={Model Structures and Fitting Criteria for System Identification with Neural Networks}, 
  year={2020},
  volume={},
  number={},
  pages={1-6}}

@article{SCHOUKENS2020310,
title = {On the Initialization of Nonlinear LFR Model Identification with the Best Linear Approximation},
journal = {IFAC-PapersOnLine},
volume = {53},
number = {2},
pages = {310-315},
year = {2020},
note = {21st IFAC World Congress},
issn = {2405-8963},
author = {Maarten Schoukens and Roland Tóth}
}

@book{Bishop1995NeuralNF,
    author = {Bishop, Christopher M},
    title = {Neural Networks for Pattern Recognition},
    publisher = {Oxford University Press},
    year = {1995},
    month = {11},
    isbn = {9780198538493},
}

@ARTICLE{DavidvandenHurk,
  author={van den Hurk, David and Weiland, Siep and van Berkel, Koos},
  journal={IEEE Transactions on Semiconductor Manufacturing}, 
  title={Control of Thermo-Mechanical Wafer Deformations in EUV Lithography Using an Active Wafer Clamp}, 
  year={2020},
  volume={33},
  number={1},
  pages={96-102}}

@article{HOEKSTRA2025101304,
title = {Learning-based model augmentation with LFRs},
journal = {European Journal of Control},
volume = {86},
pages = {101304},
year = {2025},
note = {Special Issue on the European Control Conference 2025},
issn = {0947-3580},
author = {Jan H. Hoekstra and Chris Verhoek and Roland Tóth and Maarten Schoukens}
}

@misc{Karniadakis2021,
   author = {George Em Karniadakis and Ioannis G. Kevrekidis and Lu Lu and Paris Perdikaris and Sifan Wang and Liu Yang},
   issn = {25225820},
   issue = {6},
   journal = {Nature Reviews Physics},
   month = {6},
   pages = {422-440},
   publisher = {Springer Nature},
   title = {Physics-informed machine learning},
   volume = {3},
   year = {2021}
}

@article{10.5555/70405.70408,
author = {Hornik, K. and Stinchcombe, M. and White, H.},
title = {Multilayer feedforward networks are universal approximators},
year = {1989},
issue_date = {1989},
publisher = {Elsevier Science Ltd.},
address = {GBR},
volume = {2},
number = {5},
issn = {0893-6080},
journal = {Neural Netw.},
month = jul,
pages = {359–366},
numpages = {8}
}

@article{Ruderman2011,
   author = {Michael Ruderman and Torsten Bertram},
   isbn = {9783902661937},
   issn = {1474-6670},
   issue = {1},
   journal = {IFAC Proceedings Volumes},
   month = {1},
   pages = {10764-10769},
   publisher = {Elsevier},
   title = {Modified Maxwell-slip Model of Presliding Friction},
   volume = {44},
   year = {2011}
}

@article{Qin2024,
   author = {Huibin Qin and Zefeng Zhang and Zhili Hou and Lina Li and Kai Liu and Shaoping Bai},
   issn = {2667-3797},
   issue = {3},
   journal = {Biomimetic Intelligence and Robotics},
   month = {9},
   pages = {100171},
   publisher = {Elsevier},
   title = {Modeling and analysis of hysteresis using the Maxwell-slip model for variable stiffness actuators},
   volume = {4},
   year = {2024}
}

@ARTICLE{1064347,
  author={Mayergoyz, I.},
  journal={IEEE Transactions on Magnetics}, 
  title={Mathematical models of hysteresis}, 
  year={1986},
  volume={22},
  number={5},
  pages={603-608}}

@article{SJOBERG19951691,
title = {Nonlinear black-box modeling in system identification: a unified overview},
journal = {Automatica},
volume = {31},
number = {12},
pages = {1691-1724},
year = {1995},
note = {Trends in System Identification},
issn = {0005-1098},
author = {Jonas Sjöberg and Qinghua Zhang and Lennart Ljung and Albert Benveniste and Bernard Delyon and Pierre-Yves Glorennec and Håkan Hjalmarsson and Anatoli Juditsky}
}

@article{KRIKELIS2021702,
title = {Artificial Neural Network Hysteresis Operators for the Identification of Hammerstein Hysteretic Systems},
journal = {IFAC-PapersOnLine},
volume = {54},
number = {7},
pages = {702-707},
year = {2021},
note = {19th IFAC Symposium on System Identification SYSID 2021},
issn = {2405-8963},
author = {Konstantinos Krikelis and Koos {van Berkel} and Maarten Schoukens}
}

@book{LjungGlad1994,
  author    = {Ljung, Lennart and Glad, Torkel},
  title     = {Modeling of Dynamic Systems},
  publisher = {Prentice Hall},
  address   = {Englewood Cliffs, NJ},
  year      = {1994}
}

@article{Bolderman20242,
  author  = {Bolderman, Max and Butler, Hans and Koekebakker, Sjirk and
             van Horssen, Eelco and Kamidi, Ramidin and Spaan-Burke, Theresa and
             Strijbosch, Nard and Lazar, Mircea},
  title   = {Physics-guided neural networks for feedforward control with
             input-to-state-stability guarantees},
  journal = {Control Engineering Practice},
  volume  = {145},
  pages   = {105851},
  year    = {2024}
}

@article{54e23470bab2456ebc98e4c620a3314b2,
  author  = {Veldman, Dani{\"e}l W. M. and Fey, Rob H. B. and Zwart, Hans and
             van de Wal, Marc M. J. and van den Boom, Joost D. B. J. and
             Nijmeijer, Henk},
  title   = {Optimal Thermal Actuation for Mitigation of Heat-Induced Wafer
             Deformation},
  journal = {IEEE Transactions on Control Systems Technology},
  volume  = {29},
  number  = {2},
  pages   = {514--529},
  year    = {2021}
}

@article{Schuderer20252,
  author    = {Schuderer, Matthias and Rill, Georg and Schaeffer, Thomas and
               Schulz, Carsten},
  title     = {Friction modeling from a practical point of view},
  journal   = {Multibody System Dynamics},
  volume    = {63},
  number    = {1},
  pages     = {141--158},
  year      = {2025},
  publisher = {Springer}
}
\end{document}